\documentclass[a4paper,english,preprint]{revtex4}
\usepackage{babel}
\usepackage{amsmath}
\usepackage{amssymb}
\usepackage{graphicx}

\pdfpageheight\paperheight
\pdfpagewidth\paperwidth

\newcommand{\beq}{\begin{equation}}
\newcommand{\eeq}{\end{equation}}
\newcommand{\beqs}{\begin{equation*}}
\newcommand{\eeqs}{\end{equation*}}

\begin{document}

\title{Electrostatic gyrokinetic simulation of global tokamak boundary plasma and the generation of nonlinear intermittent turbulence}

\date{\today}

\author{S. \surname{Ku}}
\email{sku@pppl.gov}
\selectlanguage{english}%
\affiliation{Princeton Plasma Physics Laboratory, Princeton University, Princeton, New Jersey 08543, USA}

\author{R.M. \surname{Churchill}}
\affiliation{Princeton Plasma Physics Laboratory, Princeton University, Princeton, New Jersey 08543, USA}

\author{C.S. \surname{Chang}}
\affiliation{Princeton Plasma Physics Laboratory, Princeton University, Princeton, New Jersey 08543, USA}

\author{R. \surname{Hager}}
\affiliation{Princeton Plasma Physics Laboratory, Princeton University, Princeton, New Jersey 08543, USA}

\author{E.S. \surname{Yoon}}
\affiliation{Rensselaer Polytech Inst, 110 8th St, Troy, NY 12180, USA}

\author{M. \surname{Adams}}
\affiliation{Lawrence Berkeley National Laboratory, 1 Cyclotron Rd, Berkeley, CA 94720, USA}

\author{E. \surname{D'Azevedo}}
\affiliation{Oak Ridge Natl Lab, POB 2008, Oak Ridge, TN 37831, USA}

\author{P. H. \surname{Worley}}
\affiliation{PHWorley Consulting, Oak Ridge, TN 37830, USA}

\begin{abstract}
Boundary plasma physics plays an important role in tokamak confinement, but is difficult to simulate in a gyrokinetic code due to the scale-inseparable nonlocal multi-physics in magnetic separatrix and  open magnetic field geometry.  Neutral particles are also an important part of the boundary plasma physics. In the present paper, novel  electrostatic gyrokinetic techniques to simulate the flux-driven, low-beta electrostatic boundary plasma is reported. Gyrokinetic ions and drift-kinetic electrons are utilized without scale-separation between the neoclassical and turbulence dynamics.  It is found that the nonlinear intermittent turbulence is a natural gyrokinetic phenomenon in the boundary plasma in the vicinity of the magnetic separatrix surface and in the scrape-off layer.   
\end{abstract}

\maketitle

\section{Introduction}

Understanding physics in the boundary region of a tokamak plasma is critically important for the fusion performance in the core plasma and for the integrity of the material surface surrounding the boundary plasma.  The global plasma confinement and the divertor heat-load width depends on the transport property of the edge plasma.  However, understanding the boundary plasma has been difficult due to the nonlinear multiscale nature of the scale-inseparable multi-physics, magnetic separatrix, plasma interaction with material wall, and the existence of neutral particles.  The scale inseparable multi-scale physics includes background profile evolution, neoclassical particle orbit dynamics, and plasma turbulence and instability.  Since the boundary plasma is in a non-thermal equilibrium state with strong sources and sinks, and the particle orbital motions sampling widely different physical regions, it is not in a Maxwellian state.  A fluid approximation may not produce a high fidelity boundary physics.  The best way to understand the boundary physics at high fidelity at this time is to use a large-scale gyrokinetic simulation.  However, the existing  gyrokinetic codes developed for the core plasma have difficulty in simulating the boundary plasma due to the difficulties described above.

The gyrokinetic code XGC1\cite{Ku2009,Ku2016} has been developed to study specifically such a boundary plasma.  The purpose of this paper is to report the novel kinetic simulation techniques that enable electrostatic simulation of the low-beta boundary plasma in contact with the material wall and across the magnetic separatrix surface using subcycled kinetic electrons.  A simulation example will be presented that produces the nonlinear intermittent turbulence that includes the so-called ``blobs'' \cite{DIppolito2011} as a natural consequence of the gyrokinetic microturbulence with kinetic electrons in the vicinity of the magnetic separatrix and in the scrape-off layer.

The paper is organized as follows:
In section \ref{sec.xgc1}, the simulation model and the novel algorithms used by XGC1 is described that can handle the non-Maxwellian tokamak edge plasmas in contact with material wall. 
In section \ref{sec.blob}, a simulation example is presented that produces the nonlinear intermittent turbulence.
Section \ref{sec.summary} gives summary and discussions.

\section{XGC1}\label{sec.xgc1}
XGC1 is a 5D gyrokinetic turbulence transport code specialized for tokamak edge simulation. 
The simulation domain includes the X-point geometry and the magnetic separatrix surface, and it is usually extended to the whole plasma including the magnetic axis and wall boundary for more proper boundary conditions. 
XGC1 can use the experimental magnetic field.   
XGC1 can solve electromagnetic perturbations, but only the electrostatic perturbation is considered in this work for low-$\beta$ boundary physics study around the magnetic separatrix and in the scrape-off layer.

\subsection{Gyrokinetic equations}
XGC1 in the electrostatic limit solves the 5D gyrokinetic Boltzmann equation\cite{Littlejohn,Hahm1988}, using the following Lagrangian equations of motion,
\begin{eqnarray}
\frac{\partial f}{\partial t} & + & \dot{\bf X} \cdot \frac{\partial f}{\partial {\bf X}} + \dot{v_\|} \cdot \frac{\partial f}{\partial v_\|} =S(f) \label{eq.Motion} \\
\dot{\bf X} & = & \frac{1}{D} \left[ v_\| \hat{b} + \frac{m  v_\|^2}{qB^2} \nabla \times \hat{b} + \frac{1}{qB^2} {\bf B} \times (\mu \nabla B -q{\bf \bar E}) \right] \nonumber \\
\dot{v_\|} & = & -\frac{1}{mD} \left( \hat{b} + \frac{m v_\|}{qB}\nabla  \times \hat{b}\right) \cdot (\mu \nabla B -q{\bf \bar E}) \nonumber \\
D & = & 1+\frac{m v_\|}{qB} \hat{b}\cdot (\nabla \times \hat{b}). \nonumber
\end{eqnarray}
Here $f$ is the distribution function of the gyrokinetic particles, ${\bf X}$ is the gyro-center position in real space,
$S(f)$ is sum of operators which is not conserving phase space volume, such as the Coulomb collisions and the heating/cooling sources. 
$v_\|$ is the velocity of the gyro-center parallel to the local magnetic field $\bf{B}$, $\hat{b}={\bf B}/B$,
$\mu=mv_\perp^2/2B$ is the magnetic moment, ${\bf \bar E}$ is the gyro-averaged electric field, $m$ is the mass, and $q$ is the charge.

The electric potential is determined by the quasi-neutrality equation. The polarization density gives the lowest order  gyrokinetic Poisson equation\cite{Hahm1988},  
\beq\label{eq.poisson1}
\nabla_\perp \cdot \frac{n_e m}{e B^2} \nabla_\perp \Phi = \bar n_i - n_e,
\eeq
where $k_\perp$ is the perpendicular wave number, $\rho_i$ is the ion gyro-radius, $n_i$ is the ion gyro-center density (not real ion density), $\bar x$ means gyro-averaging of $x$ , $n_e$ is the electron density, $\nabla_\perp$ is the perpendicular gradient operator to $\bf B$, and the ion species is assumed to have single elementary charge $e$. 
Electrons are drift kinetic in this work. 

\subsection{Hybrid-Lagrangian total $\delta f$ scheme}
To solve the gyrokinetic Boltzmann equation, XGC1 uses the hybrid-Lagrangian total-$\delta f$ scheme\cite{Ku2016}, which utilizes the phase space grid in addition to the usual marker particles.
If we define $D/Dt$ as the derivative of left hand side of Eq.~(\ref{eq.Motion}), the Boltzmann equation can be simply written as
\beq \label{eq.dfdt}
\frac{Df}{Dt}=S(f).
\eeq
Note that $D/Dt$ operator conserves phase space volume\cite{Hahm1988}.

In the $\delta f$ scheme, the distribution function $f$ is decomposed into $f = f_0 + \delta f$, and the Boltzmann equation becomes
\beq \label{eq.ddfdt}
\frac{D\delta f}{Dt}=-\frac{Df_0}{Dt} + S(f).
\eeq
Note that the above $\delta f$ equation does not use any approximation or assumption on the magnitude of $\delta f$. In some core plasmas $\delta f$ can be assumed to be small compared to the Maxwellian $f_0$, and the $D/Dt$ operator on the right-hand side can omit magnetic drifts for studying the turbulence phenomena only without the neoclassical physics. 
However, in tokamak edge plasmas, this assumption is hard to justify since it is a strong driver to non-Maxwellian distribution and the neoclassical particle dynamics is important.
Without this approximation, Eq.~(\ref{eq.ddfdt}) is mathematically identical to the original full-f (total-f) equation, Eq.~(\ref{eq.dfdt}). 
Also, it has been numerically shown that the same self-organized quasi-equilibrium state can be obtained with the total-$\delta f$ scheme as the full-f scheme does \cite{Ku2009}.

In the hybrid-Lagrangian scheme, $f$ consists of an analytic function $f_a$ (which is usually Maxwellian), $f_g$ on a 5D phase space continuum grid, and weighted particle distribution function $f_p$. 
\beq 
 f= f_a + f_g +  f_p
\eeq
In every time step, $f_g$ takes a small fraction of $f_p$, eventually leading to $f_g$ absorb the slowly varying component of $f_p$. This can reduce the magnitude of $f_p$ and the statistical noise.  

In this scheme, the total $f$ can be evaluated on the 5D continuum grid. $S(f)$ of Eq.~\ref{eq.ddfdt} is then evaluated on the 5D continuum grid and is transferred back to particle weights. Detailed algorithms and verifications are described in the reference \cite{Ku2016}.

\subsection{Kinetic electrons and subcycling}
In this study, the wave frequency $\omega$ of interest is much smaller than $k v_{e,th}$, where $k$ is the wave vector of interest and $v_{e,th}$ is the thermal speed of electrons.   
With this slow wave frequency, the electron response to the perturbed electric potential is dominated by adiabatic response on flux surface.
To reduce the particle noise from the adiabatic response, the analytic part of the electron distribution function, $f_a^e$, is chosen to be Maxwellian with the adiabatic response to perturbed potential on flux surface \cite{splitweight2000}.
\beq
f_{a}^{e} = \frac{n_0}{T_e^{3/2}} \exp\left(-\frac{K}{T_e} + \frac{e\delta\Phi}{T_e}\right),
\eeq 
where $n_0$ is equilibrium electron density, $T_e$ is the equilibrium electron temperature, and $K$ is the kinetic energy of electron, and $\delta\Phi$ is the potential deviation from flux averaged potential, $\delta \Phi \equiv \Phi - \left<\Phi\right>$. 

Since the electron density from $f_a^e$ depends on $\delta \Phi$, the gyrokinetic Poisson equation, Eq.~(\ref{eq.poisson1})  becomes
\beq\label{eq.poisson2}
\nabla_\perp  \cdot \frac{n_i m}{e B^2} \nabla_\perp \Phi  + n_0 \left[\exp\left(\frac{\delta \Phi}{T_e}\right)-1\right]= \delta \bar n_i - \delta n_e^{NA},
\eeq
where  unperturbed ion density $\bar n_{i0} = \int {\bar f}_a^i = n_0$  is assumed, non-adiabatic electron density $\delta n_e^{NA} = \int (f_p^e + f_g^e) d^3 v$, and perturbed ion density $\delta n_i = \int ( \bar f_p^i + \bar f_g^i) d^3 v$.

To increase numerical efficiency by reducing communications between parallel processors, XGC1 uses an electron sub-cycling scheme\cite{adam1982}, and the electric field is updated with ion time step, which is about $\sqrt{m_i/m_e}$ times larger than electron time step. 
This sub-cycling scheme is justified by the assumption $\omega \ll k v_{e,th}$.  

One issue in the electrostatic simulation of tokamak plasmas is the so-called $\omega_H$ mode\cite{Lee1987}.  
$\omega_H$ mode can give rapid numerical instability without self-consistent magnetic perturbations unless the time step is small enough to resolve this. 
To resolve this issue, we utilize the numerical scheme of fluid-kinetic hybrid electron model\cite{Lin2001}.
This scheme has prediction and correction phases. 
In the prediction phase, the lowest order perturbed potential, $\delta \Phi^{(0)}$ is determined by the adiabatic electron response ($\delta n_e^{NA}=0$), and $\delta \Phi^{(0)}$  is used to set $f_a^e$ in weight evolution equation. 
In the correction phase, the second order potential $\delta \Phi^{(1)}$ is obtained using the Poisson equation (\ref{eq.poisson2}) with the non-adiabatic density response $\delta n_e^{NA}$, which is from the weight evolution equation with $f_a^e (\delta \Phi^{(0)})$.
Since electron particle weights are required only when evaluating $\delta n_e^{NA}$ and XGC1 uses direct weight evolution, evaluating weight is required only at ion time step.

The calculation of electron subcycling is the most time consuming calculation and takes more than 50\% of computation time in production runs. 
XGC1 utilizing GPUs using CUDA FORTRAN together with CPUs.

\subsection{Poisson solver and field following mesh}

The nonlinear gyrokinetic Poisson equation is recently developed in XGC1, but linearized equation is used in this work.
\beq\label{eq.poisson3}
\nabla_\perp  \cdot \frac{n_0 m}{e B^2} \nabla_\perp \Phi  + n_0 \left(\frac{\delta \Phi}{T_e}\right)= \delta \bar n_i - \delta n_e^{NA},
\eeq 
Eq.~(\ref{eq.poisson3}) is converted into matrix equation using finite element method.
One numerical difficulty of the matrix equation is from the flux surface average operator in $\delta \Phi = \Phi - \left< \Phi \right >$, because the flux surface average operator appears as dense matrix. XGC1 solves the matrix equation iteratively to avoid explicit inverse of the dense matrix of flux surface average operator.  

In the drift waves we are interested in, parallel wave number $k_{||} \sim 1/R$  is much smaller than $k_\perp$. Since $E_{||}$ is important to Landau damping of waves, calculating parallel derivative of potential should be handled carefully, to avoid numerical error in $k_\perp$ calculation from larger $k_\perp$. XGC1 uses element-wise field following grid in each toroidal domain (domain between poloidal planes), and interpolation is used across the domains to get charge density and electric field in real coordinates. 
To minimize the error from interpolation, the node points of the mesh are chosen to follow field line as much as possible. Except the region near X-point, the node points falls on another node points approximately when it follows the magnetic field to the next poloidal plane.

\subsection{Wall boundary}

XGC1 includes the wall (divertor and limiter) as boundary condition of plasmas.
In the physics model, the particles are absorbed to the boundary unless reflected by the sheath potential.
In full-f method, the marker particles that represent absorbed physical particles can simply be removed from the simulation. 
In total-$\delta f$ method, those marker particles should remain in the simulation region to represent $f_p$ which describes the change from Maxwellian distribution.
Hence, we model the marker particle motion of absorbed particle so that it is reflected by infinite potential well at the boundary, which conserves the phase space volume. 
The absorption of physical particles is described by weight change, $\Delta w = - (f_a + f_g) / g - w$, where $g$ is marker particle density.  
This weight change can be very large when $f_g=0$, since $f_p$ needs to cancel $f_a$. 
When $f_g$ reaches to steady state level using the hybrid Lagrangian scheme, the weight change can be reduced to fluctuation level from time averaged $f$.

The level of sheath potential is important to determine whether an electron is reflected at the sheath or absorbed to the wall boundary. 
The size of sheath is about Debye length, which is much smaller than ion gyro-radius or the wave length we are interested in. 
XGC1 uses a modified logical sheath boundary condition\cite{Parker1993} to avoid resolving the fine structure and the high frequency ($\sim$plasma oscillation frequency) of sheath physics. 
The sheath potential is determined by the number of ions and electrons crossing the wall boundary and chosen so that the number of electrons which has larger parallel energy than the sheath potential is the same as the number of ions on average.   To avoid fast time scale oscillation, some time delay is imposed in addition to the original logical sheath algorithm.
The sheath potential $\Phi_s$ is adjusted with
\beqs
\frac{\partial \Phi_s }{\partial t} = - C (\Gamma_{i} - \Gamma_{e}),
\eeqs
where $\Gamma_{i,e}$ are the particle fluxes to the wall, $C$ is a coefficient to give time delayed ambipolar flux of gyro-center particles.


\subsection{Coulomb collision, heat source, and neutral atomic physics}

The right hand side term, $S(f)$ of Eq.~\ref{eq.Motion} includes Coulomb collision, plasma heating, radiative cooling, and neutral atomic physics. 
Since those operators do not conserve phase space volume, XGC1 does not apply those operators to particle phase variables but only to particle weight and/or $f_g$. 
Hence, the change of particle weights and $f_g$ satisfies $\Delta w g + \Delta f_g = S(f) \Delta t$ with small enough time step $\Delta t$.
Note that $g$ does not change since particle phase variables is unchanged.
$S(f)$ is evaluated in phase space grid using continuum method.
In the simulation, $\Delta w g = S(f)$ when there exist particle in the phase space grid, and $\Delta f_g = S(f)$ otherwise.

The Coulomb collision operator of $S(f)$ is solved using a fully nonlinear Fokker-Plank-Landau collision operator on the 2D velocity space grid\cite{Yoon2014,Hager2016}. 
The nonlinear Coulomb collision operator is necessary especially due to the non-Maxwellian distribution from steep pedestal gradient and absorbed particles to the wall boundary.

Heating and radiative cooling are applied in an isotropic way in velocity space with small adjustment to conserve net parallel momentum.
The radial profiles of heating and cooling are from prescribed profiles and/or from simple radiative power loss model using neutral and impurity profiles.

XGC1 has built-in Monte-Carlo neutral profile simulator consistent with plasma profile evolution.
Using the neutral profile, the ionization of neutral particles and charge exchange between neutral and ions are calculated using the rate coefficients of atomic processes.


\section{An example simulation of boundary plasma}\label{sec.blob}
\subsection{Simulation setup}

\begin{figure}[h]
  \centering
  \mbox{\includegraphics[scale=0.6]{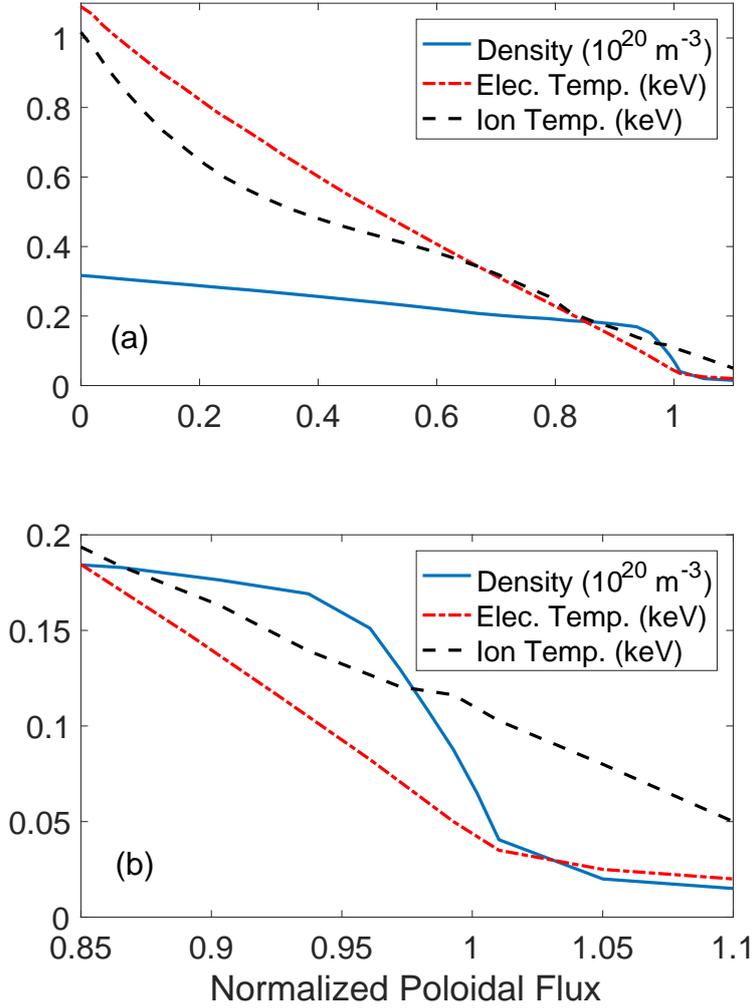}}
  \vspace{-40pt}
  \caption{The initial plasma density, electron temperature, and ion temperature profiles of (a) whole simulation domain and (b) edge region. }
  \label{fig.profiles}
\end{figure}

In the example simulation, we used a model based on a DIII-D like L-mode plasma profiles using the magnetic equilibrium 146598 at 1105ms.
The model profiles of plasma density and temperature are shown in Fig.~\ref{fig.profiles}.  
At the separatrix the plasma density is $7\times 10^{17} m^{-3}$ and the electron temperature is 45 eV. This gives 0.003\% of plasma beta, and the electrostatic turbulence simulation can be justified.


The simulation is performed on the Titan Cray-XK7 at the Oak Ridge Leadership Computing Facility using 16,384 computing nodes (88\% of Titan) for 72 hours.
The number of marker particles is 13 billion for each plasma species.
The number of node points of the real space triangular mesh is about 55,000 per poloidal plane (toroidal cross section).
32 poloidal planes are used.
The velocity space mesh is 30 by 31 rectangular grid with the maximum $v_{||}$ and $v_{\perp}$ being 3 times the thermal velocity.
Hence, the total phase space grid has about 1.5 billion grid points.
The simulation time step is  $ 1.6 \times10^{-7}$ sec, and the total simulation time is 1.7 ms,
until the turbulence in the boundary plasma reaches a steady state. 
The steady state heat transport is also achieved only in the edge region at the end of the simulation.
Coulomb collision is not applied to speed up the simulation in this example problem, even though it is normally applied in other physics simulations\cite{ChangIAEA2016}.

\begin{figure}[h]
  \centering
  \mbox{\includegraphics[scale=0.7]{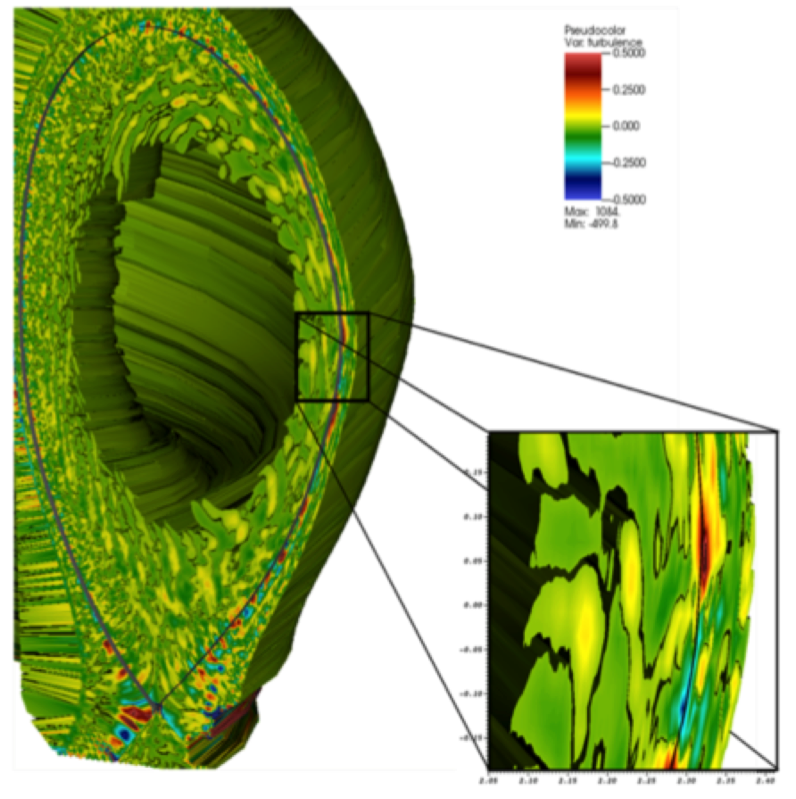}}
  \caption{Poloidal cross-sectional view of the perturbed density from electrostatic turbulence in the global boundary plasma.}
  \label{fig.pol_cross}
\end{figure}

\subsection{Observation of nonlinear turbulence}

Figure~\ref{fig.pol_cross} is the poloidal cross-sectional view of the nonlinear electrostatic turbulence in the global boundary plasma.
Insert box shows enlarged structure of the turbulence. 
Streamer type structures can be seen inside the magnetic separatrix (ion temperature gradient turbulence and trapped electron turbulence). 
Around the magnetic separatrix and in the scrape-off layer, blobby type structures can be seen.

\begin{figure}[h]
  \centering
  \mbox{\includegraphics[scale=0.5]{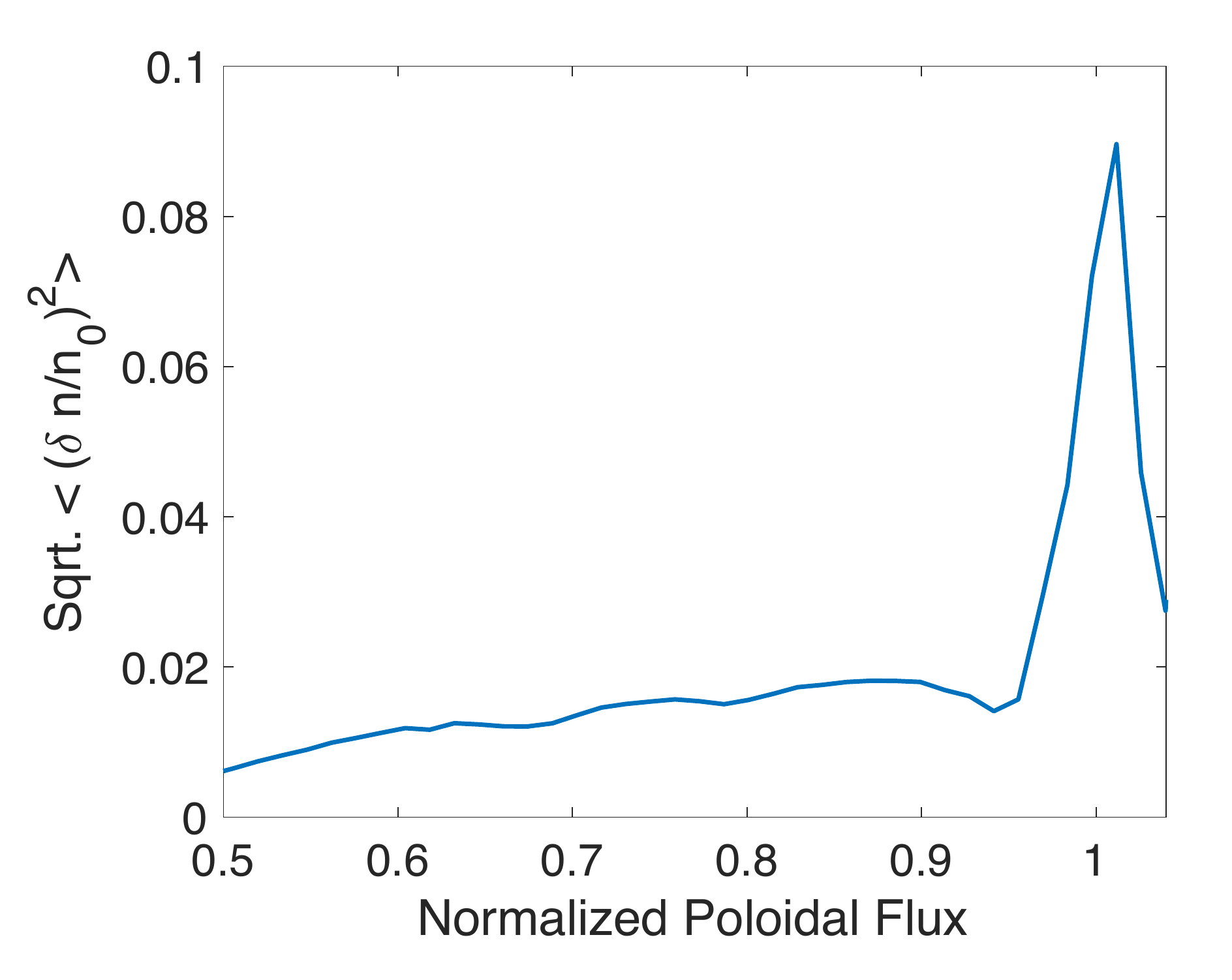}}
  \caption{Square root of turbulence intensity of normalized perturbed density measured at the outside midplane, averaged over $-30$ cm to 30 cm in poloidal direction, 0 to $2 \pi$ in toroidal direction, and 0.6 - 1.7 ms in time}
  \label{fig.turb_inten}
\end{figure}

Figure~\ref{fig.turb_inten} shows square root of space-time averaged turbulence intensity of normalized perturbed density $\delta n / n_0$ measured at the outside midplane, where 
$\delta n$ is the perturbation from toroidally averaged density and $n_0$ is the equilibrium density.
$(\delta n/n_0)^2$ is averaged over $-30$ cm to 30 cm in the poloidal direction, 0 to $2 \pi$ in the toroidal direction, and 0.6 - 1.7 ms in time, when turbulence intensity is nonlinearly saturated. 
In the average operation, flux surface volume is weighted.
The turbulence intensity is rapidly increasing from $\psi_n = 0.96$ where the large density gradient starts. The turbulence intensity  peaks around the separatrix. 
The blob activity also peaks in this region.

\begin{figure}[h]
  \centering
  \mbox{\includegraphics[scale=0.4]{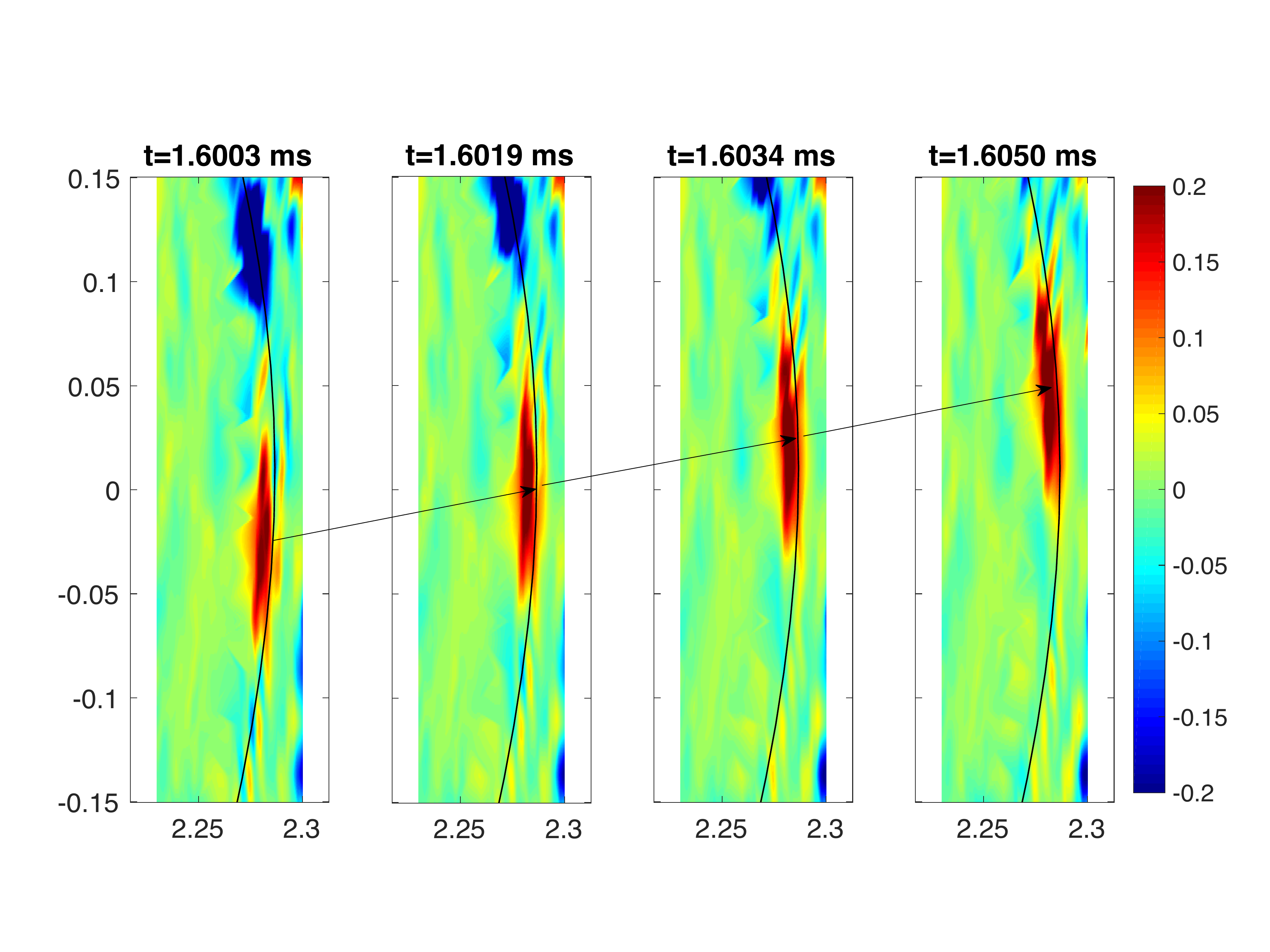}}
  \caption{Normalized density perturbation from toroidal average at time of 1.6003 ms, 1.6019 ms,  1.6034 ms and 1.6050 ms. 
  The black lines represent magnetic separatrix. 
  The arrows indicate the motion of the center of the blob.
  }
  \label{fig.blobs}
\end{figure}

Figure~\ref{fig.blobs} shows a blob movement observed in the edge region of outside midplane.
The colormap of the figure represents normalized density perturbation from toroidal average.
Each sub-figure has time slice of 1.6003 ms, 1.6019 ms,  1.6034 ms and 1.6050 ms, from left to right.
The blobs are located around separatrix.
The size of blobs are about 7 cm poloidally and 1.5 cm radially. 
The blobs show about 10 - 20\% density perturbations.

\begin{figure}[h]
  \centering
  \mbox{\includegraphics[scale=0.6]{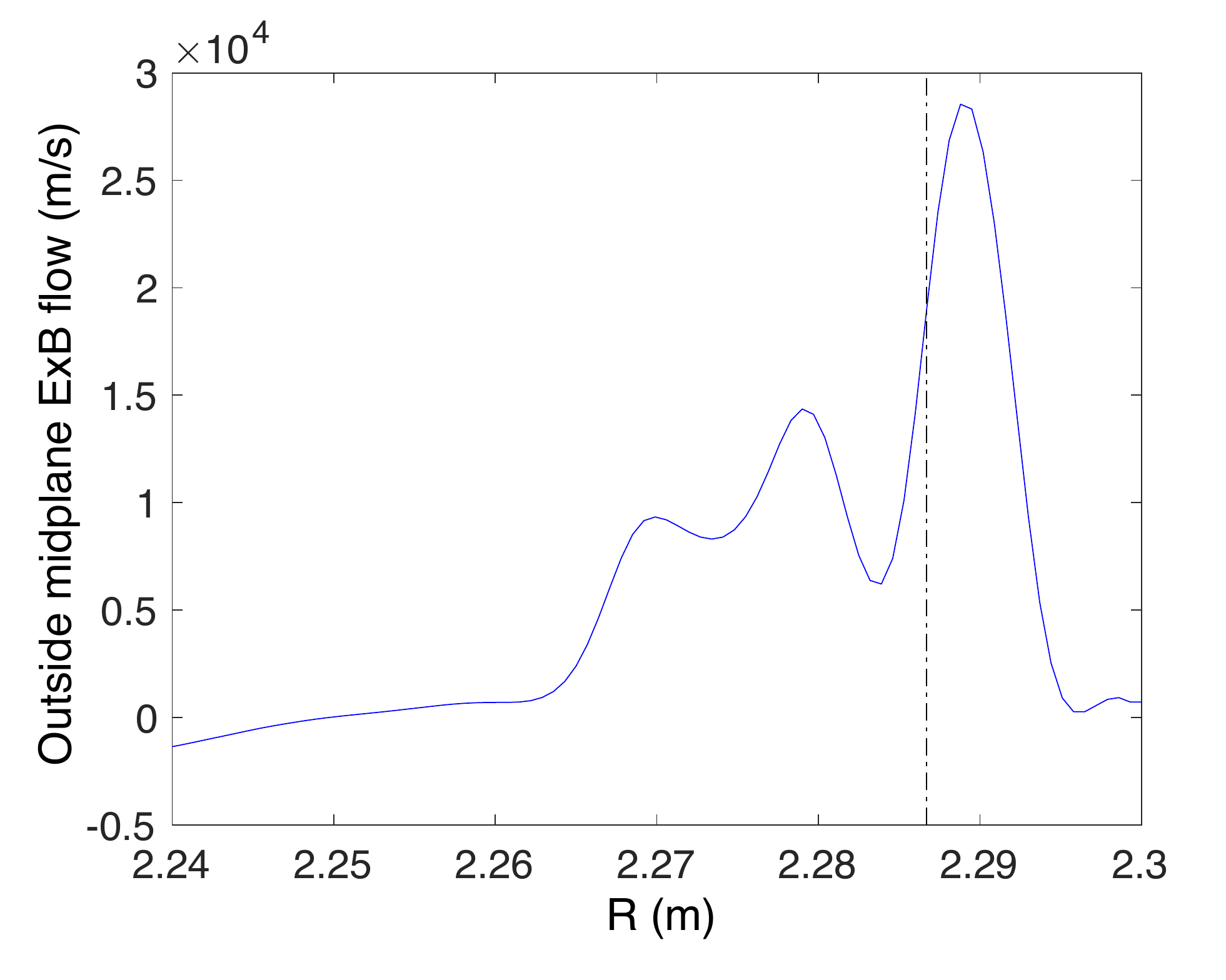}}
  \caption{Poloidal ExB flow at the outside midplane when t = 1.60 ms. 
  The dashed vertical line is the location of separatrix. 
  Tracking of the blobs near the separatrix shows poloidal speed near 15 km/s, close to the ExB speed.
  }
  \label{fig.exb}
\end{figure}

Tracking the center of the blob of the figure, the poloidal speed of the blob is about 15 km/s.
The poloidal speed is close to the poloidal ExB flow at the location of the blob, which is shown in Fig.~\ref{fig.exb}.
In this example with a rather strong density pedestal, the poloidal ExB speed is strong, and the poloidal motion of the blob is mostly given by the poloidal ExB flow.
The radial motion of the blobs is much smaller than poloidal motion at this location.
In this case with strong ExB flow, blobs move out rapidly from inside the separatrix to the scrape-off layer well above the midplane.
A systematic analysis of blob dynamics will be studied in a follow-up paper.

\section{Summary}\label{sec.summary}

Boundary plasma physics plays an important role in tokamak confinement, but is difficult to simulate in a gyrokinetic code due to the scale-inseparable nonlocal multi-physics in magnetic separatrix and  open magnetic field geometry.  
Neutral particles are also an important part of the boundary plasma physics. 
In the present paper, novel  electrostatic gyrokinetic techniques to simulate the flux-driven, low-beta electrostatic boundary plasma is reported. Gyrokinetic ions and drift-kinetic electrons are utilized without scale-separation between the neoclassical and turbulence dynamics.  
It is found that the nonlinear intermittent turbulence is a natural gyrokinetic phenomenon in the boundary plasma in the vicinity of the magnetic separatrix surface and in the scrape-off layer.

 
\section*{Acknowledgement}
This work is supported by the U.S. Department of Energy under contract No. DE- AC02-09CH11466. 
This research used resources of the Oak Ridge Leadership Computing Facility at the Oak Ridge National Laboratory, which is supported by the Office of Science of the U.S. Department of Energy under Contract No. DE-AC05-00OR22725.

\bibliography{kinetic_elec}

\begin{thebibliography}{13}
\expandafter\ifx\csname natexlab\endcsname\relax\def\natexlab#1{#1}\fi
\expandafter\ifx\csname bibnamefont\endcsname\relax
  \def\bibnamefont#1{#1}\fi
\expandafter\ifx\csname bibfnamefont\endcsname\relax
  \def\bibfnamefont#1{#1}\fi
\expandafter\ifx\csname citenamefont\endcsname\relax
  \def\citenamefont#1{#1}\fi
\expandafter\ifx\csname url\endcsname\relax
  \def\url#1{\texttt{#1}}\fi
\expandafter\ifx\csname urlprefix\endcsname\relax\def\urlprefix{URL }\fi
\providecommand{\bibinfo}[2]{#2}
\providecommand{\eprint}[2][]{\url{#2}}

\bibitem[{\citenamefont{Ku et~al.}(2009)\citenamefont{Ku, Chang, and
  Diamond}}]{Ku2009}
\bibinfo{author}{\bibfnamefont{S.}~\bibnamefont{Ku}},
  \bibinfo{author}{\bibfnamefont{C.}~\bibnamefont{Chang}}, \bibnamefont{and}
  \bibinfo{author}{\bibfnamefont{P.}~\bibnamefont{Diamond}},
  \bibinfo{journal}{Nucl. Fusion} \textbf{\bibinfo{volume}{49}},
  \bibinfo{pages}{115021} (\bibinfo{year}{2009}).

\bibitem[{\citenamefont{Ku et~al.}(2016)\citenamefont{Ku, Hager, Chang, Kwon,
  and Parker}}]{Ku2016}
\bibinfo{author}{\bibfnamefont{S.}~\bibnamefont{Ku}},
  \bibinfo{author}{\bibfnamefont{R.}~\bibnamefont{Hager}},
  \bibinfo{author}{\bibfnamefont{C.}~\bibnamefont{Chang}},
  \bibinfo{author}{\bibfnamefont{J.}~\bibnamefont{Kwon}}, \bibnamefont{and}
  \bibinfo{author}{\bibfnamefont{S.}~\bibnamefont{Parker}},
  \bibinfo{journal}{J. Comp. Physics} \textbf{\bibinfo{volume}{315}},
  \bibinfo{pages}{467} (\bibinfo{year}{2016}).

\bibitem[{\citenamefont{D'Ippolito et~al.}(2011)\citenamefont{D'Ippolito, Myra,
  and Zweben}}]{DIppolito2011}
\bibinfo{author}{\bibfnamefont{D.}~\bibnamefont{D'Ippolito}},
  \bibinfo{author}{\bibfnamefont{J.}~\bibnamefont{Myra}}, \bibnamefont{and}
  \bibinfo{author}{\bibfnamefont{S.}~\bibnamefont{Zweben}},
  \bibinfo{journal}{Phys. Plasmas} \textbf{\bibinfo{volume}{18}},
  \bibinfo{pages}{060501} (\bibinfo{year}{2011}).

\bibitem[{\citenamefont{Littlejohn}(1985)}]{Littlejohn}
\bibinfo{author}{\bibfnamefont{R.~G.} \bibnamefont{Littlejohn}},
  \bibinfo{journal}{Phys. Fluids} \textbf{\bibinfo{volume}{28}},
  \bibinfo{pages}{2015} (\bibinfo{year}{1985}).

\bibitem[{\citenamefont{Hahm}(1988)}]{Hahm1988}
\bibinfo{author}{\bibfnamefont{T.~S.} \bibnamefont{Hahm}},
  \bibinfo{journal}{Phys. Fluids} \textbf{\bibinfo{volume}{31}},
  \bibinfo{pages}{2670} (\bibinfo{year}{1988}).

\bibitem[{\citenamefont{Manuilskiy and Lee}(2000)}]{splitweight2000}
\bibinfo{author}{\bibfnamefont{I.}~\bibnamefont{Manuilskiy}} \bibnamefont{and}
  \bibinfo{author}{\bibfnamefont{W.~W.} \bibnamefont{Lee}},
  \bibinfo{journal}{Phys. Plasmas} \textbf{\bibinfo{volume}{7}},
  \bibinfo{pages}{1381} (\bibinfo{year}{2000}).

\bibitem[{\citenamefont{Adam et~al.}(1982)\citenamefont{Adam, Courdin, and
  Langdon}}]{adam1982}
\bibinfo{author}{\bibfnamefont{J.}~\bibnamefont{Adam}},
  \bibinfo{author}{\bibfnamefont{A.}~\bibnamefont{Courdin}}, \bibnamefont{and}
  \bibinfo{author}{\bibfnamefont{A.}~\bibnamefont{Langdon}},
  \bibinfo{journal}{J. Comp. Physics} \textbf{\bibinfo{volume}{47}},
  \bibinfo{pages}{229} (\bibinfo{year}{1982}).

\bibitem[{\citenamefont{Lee}(1987)}]{Lee1987}
\bibinfo{author}{\bibfnamefont{W.}~\bibnamefont{Lee}}, \bibinfo{journal}{J.
  Comput. Phys.} \textbf{\bibinfo{volume}{72}}, \bibinfo{pages}{243}
  (\bibinfo{year}{1987}).

\bibitem[{\citenamefont{Lin and Chen}(2001)}]{Lin2001}
\bibinfo{author}{\bibfnamefont{Z.}~\bibnamefont{Lin}} \bibnamefont{and}
  \bibinfo{author}{\bibfnamefont{L.}~\bibnamefont{Chen}},
  \bibinfo{journal}{Phys. Plasmas} \textbf{\bibinfo{volume}{8}},
  \bibinfo{pages}{1447} (\bibinfo{year}{2001}).

\bibitem[{\citenamefont{Parker et~al.}(1993)\citenamefont{Parker, Procassini,
  Birdsall, and Cohen}}]{Parker1993}
\bibinfo{author}{\bibfnamefont{S.}~\bibnamefont{Parker}},
  \bibinfo{author}{\bibfnamefont{R.}~\bibnamefont{Procassini}},
  \bibinfo{author}{\bibfnamefont{C.}~\bibnamefont{Birdsall}}, \bibnamefont{and}
  \bibinfo{author}{\bibfnamefont{B.}~\bibnamefont{Cohen}}, \bibinfo{journal}{J.
  Comp. Physics} \textbf{\bibinfo{volume}{104}}, \bibinfo{pages}{41}
  (\bibinfo{year}{1993}).

\bibitem[{\citenamefont{Yoon and Chang}(2014)}]{Yoon2014}
\bibinfo{author}{\bibfnamefont{E.}~\bibnamefont{Yoon}} \bibnamefont{and}
  \bibinfo{author}{\bibfnamefont{C.}~\bibnamefont{Chang}},
  \bibinfo{journal}{Phys. Plasmas} \textbf{\bibinfo{volume}{21}},
  \bibinfo{pages}{032503} (\bibinfo{year}{2014}).

\bibitem[{\citenamefont{Hager et~al.}(2016)\citenamefont{Hager, Yoon, Ku, and
  et~al.}}]{Hager2016}
\bibinfo{author}{\bibfnamefont{R.}~\bibnamefont{Hager}},
  \bibinfo{author}{\bibfnamefont{E.}~\bibnamefont{Yoon}},
  \bibinfo{author}{\bibfnamefont{S.}~\bibnamefont{Ku}}, \bibnamefont{and}
  \bibinfo{author}{\bibnamefont{et~al.}}, \bibinfo{journal}{J. Comput. Physics}
  \textbf{\bibinfo{volume}{315}}, \bibinfo{pages}{644} (\bibinfo{year}{2016}).

\bibitem[{\citenamefont{Chang et~al.}(2016)\citenamefont{Chang, Ku, and
  et.al.}}]{ChangIAEA2016}
\bibinfo{author}{\bibfnamefont{C.}~\bibnamefont{Chang}},
  \bibinfo{author}{\bibfnamefont{S.}~\bibnamefont{Ku}}, \bibnamefont{and}
  \bibinfo{author}{\bibnamefont{et.al.}}, \bibinfo{journal}{Nucl. Fusion
  Submitted} \textbf{\bibinfo{volume}{''Gyrokinetic projection of the divertor
  heat-flux width from present tokamaks to ITER"}} (\bibinfo{year}{2016}).

\end{thebibliography}

\end{document}